# Advanced chemical state studies of oxide films by lab-based HAXPES combining soft and hard X-ray sources

S. Siol[1*], J. Mann[2], J. Newman[2], T. Miyayama[3], K. Watanabe[3], P. Schmutz[1], C. Cancellieri[1], L. P. H. Jeurgens[1*]

[1] Empa, Swiss Federal Laboratories for Materials Science and Technology,
Laboratory for Joining Technologies and Corrosion, Dübendorf. Switzerland

[2] Physical Electronics, 18725 Lake Drive East, Chanhassen, MN 55317, USA

[3] Ulvac-PHI, 2500 Hagisono, Chigasaki, Kanagawa, 253-8522, Japan

**Graphical abstract:**

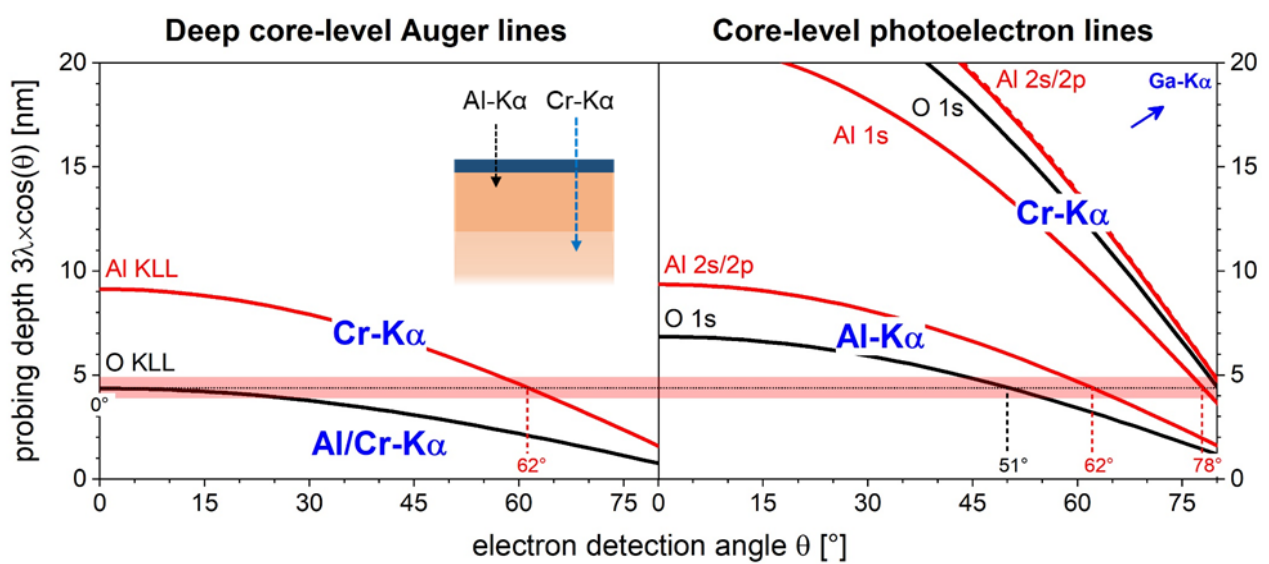

**Abstract:**

The greater information depth provided in Hard X-ray Photoelectron Spectroscopy (HAXPES) enables non-destructive analyses of the chemistry and electronic structure of buried interfaces. Moreover, for industrially relevant elements like Al, Si and Ti, the combined access to the Al 1s, Si 1s or Ti 1s photoelectron line and its associated Al KLL, Si KLL or Ti KLL Auger transition, as required for local chemical state analysis on the basis of the Auger parameter, is only possible with hard X-rays. Until now, such photoemission studies were only possible at synchrotron facilities. Recently however, the first commercial XPS/HAXPES systems, equipped with both soft and hard X-ray sources, have entered the market, providing unique opportunities for monitoring the local chemical state of all constituent ions in functional oxides at different probing depths, in a routine laboratory environment. Bulk-sensitive shallow core-levels can be excited using either the hard or soft X-ray source, whereas more surface-sensitive deep core-level photoelectron lines and associated Auger transitions can be measured using the hard X-ray source. As demonstrated for thin $Al_2O_3$, $SiO_2$ and $TiO_2$ films, the local chemical state of the constituting ions in the oxide may even be probed at near constant probing depth by careful selection of sets of photoelectron and Auger lines, as excited with the combined soft and hard X-ray sources. We highlight the potential of lab-based HAXPES for the research on functional oxides and also discuss relevant technical details regarding the calibration of the kinetic binding energy scale.

## 1. Introduction

X-ray photoelectron spectroscopy (XPS) is a powerful tool for chemical state analysis. It was established by Kai Siegbahn in the 1950s, who was awarded the Nobel Prize for his work in 1981.[1] Today XPS has become a routine characterization technique for surface and interface analysis in many laboratories around the globe. Most commercial XPS systems that are currently on the market are based on monochromated Al-K$\alpha$ (1486.7 eV) radiation. However, many of the early XPS systems were using different anode materials, in particular, non-monochromatic Mg-K$\alpha$ (1253,6 eV) sources, as well as other non-monochromatic X-ray sources with much higher excitation energies, such as Mo-K$\alpha$ (17.479 keV)[2], Ag-L$\alpha$ (2984.3 eV)[3], Rh-L$\alpha$ (2696.8 eV)[3], Zr-L$\alpha$ (2042.4 eV).[4] Even monochromated Al-K$\alpha$ and Cr-K$\beta$ sources were combined, since the energy of the fourth order reflection of the relatively low intensity Cr-K$\beta$ line (5946.7 eV) almost exactly matches that of the Al-K$\alpha$ line (1486.6 eV).[5,6] XPS using X-rays with energies in excess of 2 keV is commonly referred to as Hard X-ray or High-Energy XPS, further denoted as HAXPES. The first HAXPES measurements, using non-monochromatic Mo-K$\alpha$ radiation, were reported as early as 1957[2] by Siegbahn's group. It is generally acknowledged that the use of hard X-rays (instead of soft X-rays) offers many advantages for dedicated XPS studies, such as:

- An increased probing depth of photoelectrons from shallow core levels as well as the valence band structure, thus allowing non-destructive analysis of the chemistry and electronic structure of thin films and their buried interfaces up to depths of about 20 nm: see Fig. 1a.
- Access to deep core-level photoelectron lines and deep-core-level Auger transitions, which are not accessible by conventional XPS, allowing advanced chemical-state studies.
- The capability to separate commonly overlapping photoelectron and Auger lines by employing different photon energies, thereby facilitating chemical-state and quantitative XPS analysis of complex multi-element compounds.
- Improved capabilities for non-destructive XPS analysis by resolving the intrinsic (primary zero-loss) and extrinsic (inelastically scattered) intensities[7,8] of multiple shallow and deep core-level photoelectron lines with their different information depths and inelastic background shapes.[9–11]

In recent days, modern synchrotron facilities provide monochromatic X-ray beams with a tuneable energy ranging from the soft X-ray (hundreds of eV, or even lower) to the hard X-ray regime (exceeding 2 keV), depending on the synchrotron design. Moreover, synchrotron X-ray beams are much more brilliant (i.e. providing higher photon fluxes) as compared to conventional lab-based X-ray sources, which is very beneficial to counter the drastic decrease of the photoionization cross section of the shallow core levels with increasing excitation energy.[12] On the downside, the very bright X-ray beam at the synchrotron can easily induce unwanted degradation in volatile and metastable materials, even for short irradiation times. In recent years, HAXPES at the synchrotron has been very successfully applied to characterize buried interfaces and bulk-like chemical and electronic structures of thin-film devices, catalysts and other functional materials.[13–15] However, comprehensive chemical state studies by HAXPES at the synchrotron in dependence of the synthesis, processing and/or environmental exposure conditions are rare,[16] which can be attributed to limited access and measurement time at synchrotron facilities and the resulting lack of a direct feedback loop between synthesis and characterization.

In recent years, the first commercial lab-based HAXPES spectrometers became available on the market[17]. At present, the available configurations typically include an Ag-Lα, Cr-Kα or Ga-Kα hard X-ray source, which can be combined with a soft Al-Kα X-ray source (further referred to as dual-source HAXPES or DS-XPS). The relatively weak signal intensity of the more bulk-sensitive, shallow core-level photoelectron lines, as excited by hard X-rays, can be compensated by employing a relatively large (e.g. spot size ≥ 100 μm) and grazing (e.g. 70° with respect to surface normal) incident X-ray beam, as well as a relatively large angular opening of the analyser lens (e.g. 20° – 30°). In addition, the pass energy for photoelectron detection can be increased, at the cost of the energy resolution. The full potential of a DS-XPS approach, analogous to conventional XPS, can be accessed by implementing the spectrometer in an integrated system with combined synthesis, processing and analysis. The strength of such an in-situ experimental approach combining DS-XPS analysis with controlled environmental processing, as will be implemented at the Empa laboratories in 2020, is the high versatility of scientific studies that can be performed on a day-to-day basis in parallel to reveal the evolution of the chemistry and electronic structure of buried interfaces in thin films, catalysts and other types of functional (nano-)materials during successive synthesis, processing and environmental exposure steps.

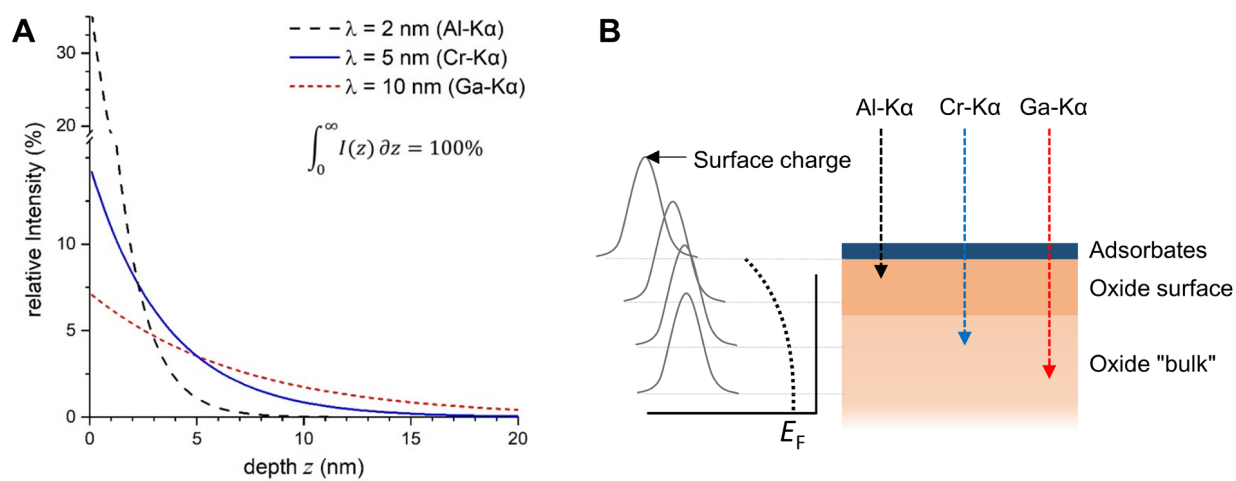

**Figure 1: A)** Normalized intensity of zero-loss photoelectrons, as detected from atoms at depth $z$ beneath the surface in a homogenous inorganic solid at θ = 45°. The trend is shown for typical values of the IMFP of photoelectrons emitted using different X-ray sources (Al-Kα → λ = 2 nm, Cr-Kα → λ = 5 nm and Ga-Kα → λ = 10 nm) **B)** The surface region of oxides is typically not homogeneous. In addition, charging of the surface can lead to depth-dependent peak shifts.

## 2.   Chemical-state analysis by lab-based HAXPES combining soft and hard X-ray sources

One major advantage of HAXPES compared to conventional XPS is the increase in information depth (see Fig. 1a), which enables non-destructive analysis of the chemistry and electronic structure at buried interfaces in functional thin films and nano-multilayer stacks, relevant for many existing and emerging nanotechnologies. For example, the different valence state of an element at a buried interface can be distinguished based on the chemical shift of the respective core-level photoelectron line with respect to its bulk or reference state.[18] Analogously, bulk-like chemical and electronic properties of functional materials with deviating surface properties (e.g. passivated metal and alloy surfaces, as well as oxidized and reduced catalysts) become accessible without detrimental sputtering effects. Notably, the increased probing depth of HAXPES can also effectively aid to mitigate daily problematics in non-destructive XPS analysis of air-exposed specimens due to inevitable surface contaminants, like adventitious carbon and/or oxygen (see Fig. 2a). Depth-resolved analysis by HAXPES at the synchrotron or by DS-XPS in the lab also provides extended capabilities for studying the electronic properties of energy conversion materials; for example, of band bending effects due to the presence of interface or surface states[19–22] and/or gradients in doping, which can cause a shift in the Fermi level (and thus of the photoelectron lines) as function of depth below the surface (see Figure 1b). Provided that the angular lens opening (or segment) for electron detection is not too large (presumably ≤ ±10°), depth-dependent gradients in valence states and/or chemical composition[18,23–25] could also be non-destructively assessed in the laboratory by combining DS-XPS with angle-resolved analysis.

The use of hard X-rays in HAXPES also provides access to a wealth of deep core-level photoelectron lines and their associated deep core level Auger transitions, which are not available by soft X-rays. The chemical shifts of Auger lines are typically larger than those of the respective photoelectron lines, which is a direct consequence of the smaller final state relaxation energy for the single core-hole final state in the photoemission process as compared with the double core-hole final state in the Auger process.[26] As first recognized by Wagner,[27,28] detailed information about the local chemical state of a given element in a compound can thus be derived from the combined analysis of its core-level photoelectron lines and the corresponding Auger electron line. The local chemical state of an element in a solid primarily comprises the first coordination sphere around the core-ionized atom, as well as the angles and distances between the core-ionized atom and its first-neighboring atoms (or ions). The modified Auger parameter (AP) $\alpha'$ of an element in a compound, as originally proposed by Wagner,[27,28] is defined as the sum of the binding energy $E_B$ of a strong core-level photoelectron line (PE) and the kinetic energy $E_K$ of a respective prominent and sharp core-level Auger transition (AE), i.e. (see also Fig. 2a)

$$\alpha' = E_K(AE) + E_B(PE) = E_K(AE) + h\nu - E_K(PE), \tag{1}$$

where $h\nu$ is the X-ray excitation energy and $E_K(PE)$ corresponds to the kinetic energy of the emitted photoelectron. The change in $E_K(PE)$ for the atom in a given compound with respect to its reference state is typically referred to as the chemical shift. The magnitude of the chemical shift, $\Delta E_K(PE)$, is the result of the differences in initial state $\Delta\varepsilon$ and final state $\Delta R^{ea}$ effects for the *single* core-ionized atom in the compound with respect to a given reference state and can be expressed by

$$\Delta E_K(PE) = -\Delta\varepsilon + \Delta R^{ea}, \tag{2}$$

where $\Delta\varepsilon$ is the energy difference of the core-level electron shell in the ground state of the atom[26,27] and $\Delta R^{ea}$ is the difference in the extra atomic relaxation (or polarization) energy for the single core-ionized atom.[26,29,30] The magnitude of $R^{ea}$ arises from the final-state screening of the core-hole state created in the photoemission process by electrons from neighboring atoms in the compound, as illustrated in Fig. 2b (i.e. for a single isolated atom: $R^{ea} \equiv 0$). Hence the value of $\Delta R^{ea}$ can be directly related to the electronic polarizability of the neighboring atoms (ligands) around the central (i.e., core-ionized) atom upon core-hole formation,[26,27] which is very sensitive to structural changes in the nearest coordination sphere of the core-ionized atom.

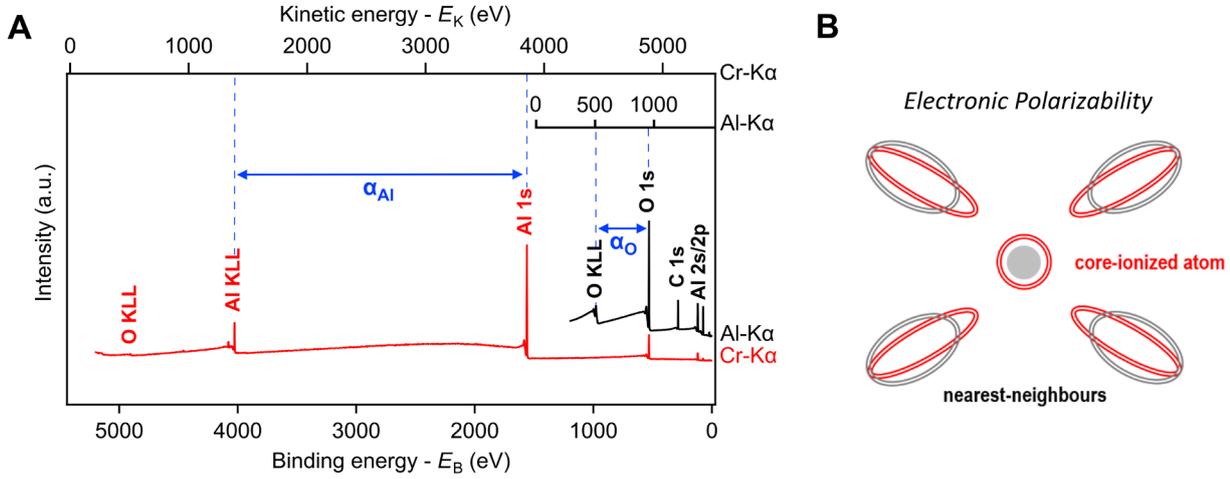

**Figure 2**: The Auger parameter is defined as the distance between the kinetic energies of a core-level photoelectron line and its deep core-level Auger electron for a given element (A) in the compound. The Auger parameter is a powerful tool for investigations of the local chemical state, as well as other important materials properties such as the electronic polarizability (B). Please note that the adventitious carbon peak, which is clearly visible in the soft Al-K$\alpha$ spectrum, has practically disappeared in the hard Cr-K$\alpha$ spectrum recorded from the identical sample.

The kinetic energy shift of a core-core Auger line $\Delta E_K$ is the result of the differences in initial state and final state effects for the *double* core-ionized atom in the compound with respect to the reference state and can be approximated by:

$$\Delta E_K (AE) \approx -\Delta\varepsilon + 3\Delta R^{ea} \qquad (3)$$

As follows from Eq. (1), subtraction of Eqs. 2 and 3 results in the AP shift with respect to its reference state

$$\Delta\alpha' \approx 2 \times \Delta R^{ea}. \qquad (4)$$

This implies that the AP shift for an atom in a given compound with respect to its reference state is proportional to the difference in extra atomic relaxation energy, which can be related to important physical properties, such as the electronic polarizability, the dielectric constant or the relative density.[31–33] By analyzing the kinetic energy difference between a core-level photoelectron and its corresponding Auger line (rather than an absolute binding energy position of a single photoelectron line), uncertainties in the position of the Fermi level, as well as the complex balance of initial and final state contributions to binding energy shifts, can be avoided. Notably, the AP value is independent of static charging and does not even require a precise calibration of the spectrometer work function, which is particularly useful for establishing changes in the chemical environment of insulating and photoactive materials.

The KLL and LMM Auger transitions are among the most intense and commonly studied Auger transitions and require the excitation of a photoelectron from core shells with quantum number n=1 (K) and n=2 (L) for triggering the associated Auger transition. In this regard, it should be emphasized that the derivation of Eq. (4) is only valid for core-core Auger transitions; i.e. KLL and LMM Auger transitions with a final core hole in L and M valence shells, respectively, should not be used. This very much restricts the number of elements available for AP studies based on relatively sharp and intense KLL and LMM Auger transitions, in particular, when using soft X-ray sources.[26,29,34] For example, the KLL transition requires the activation of the 1s photoemission process, which for monochromatic Al-K$\alpha$ radiation is only possible for elements with a mass unit of up to 12 (Mg). Consequently, elements such Al, Mg, Si, and Ti, which are crucial for advancing a broad range of existing and emerging technologies, cannot be assessed by modern XPS systems (using monochromatic Al-K$\alpha$ sources). In the past, this physical limitation was partially circumvented by exploiting the Bremsstrahlung radiation of non-monochromated X-ray sources to excite deep core-level photoelectrons for triggering the associated deep core-core Auger transition.[29,33,35–38] Bremsstrahlung generates a continuous spectrum of X-rays with an energy up to the acceleration voltage of the electrons (typically tens to hundreds of keV), which can activate the deep 1s core-level photoemission process for measuring the sharp and intense KLL Auger lines (at apparent negative binding energies). In this way, for example, the changes in the local chemical states of Al cations and O anions in $Al_2O_3$ thin films could be monitored across the amorphous-to-crystalline transition.[32] However, unfocussed non-monochromatic X-ray sources have become obsolete in modern XPS systems, since they typically rely on focussed scanning X-ray beams (with ever-smaller beam widths) and relatively wide angular openings of the hemispherical analyser input lens. Consequently, despite its indisputable scientific merit, Auger parameter analysis has become a rare sight in recent literature featuring surface and interface analysis using XPS. With the renewed interest in lab-based HAXPES, we expect a revival in chemical state studies on the basis of the Auger parameter concept.

The higher the energy of the applied X-ray excitation source, the higher the kinetic energy of the detected core-level photoelectrons. On the contrary, detected core-core Auger electrons have a fixed kinetic energy, independent of the X-ray excitation energy. This implies that (see Fig. 1a), the higher the energy of the applied X-ray source in HAXPES, the higher the probing depth of the measured core-level photoelectron line as compared to the constant probing depth of the core-core Auger electron line. This is illustrated in Fig. 3 for the angle-dependent probing depths of the Al 2s, Al 2p and O 1s core level emission lines, as well as the corresponding Al KLL and O KLL Auger electron lines, in am-$Al_2O_3$ for soft Al-K$\alpha$ and hard Cr-K$\alpha$ X-rays. According to Eq. 2, the modified APs for $Al_2O_3$ are calculated as (see Fig. 2a):

AP of O anions: $\quad \alpha'_O = E_K(\text{O KLL}) + E_B(\text{O1s})$ \hfill (5a)

AP of Al cations: $\quad \alpha'_{Al} = E_K(\text{Al KLL}) + E_B(\text{Al 1s/2s/2p})$ \hfill (5b)

As reflected in Fig. 3, when using only hard X-ray radiation (e.g. Cr-K$\alpha$ or Ga-K$\alpha$), the Auger (O KLL and Al KLL) and photoelectron lines (O1s and Al 2p) pertaining to Eqs. 5a and 5b will originate from different depths. For example, for a normal detection angle of 0° with respect to the surface normal, the differences in probing depths of the O 1s, Al 2s and Al 2p photoelectron lines for the Al-K$\alpha$ and Cr-K$\alpha$ radiation exceed 20 nm. Similar graphs of the angle-dependent probing depths of the 1s, 2s and 2p photoelectron lines and associated KLL Auger lines in $SiO_2$ and $TiO_2$ for soft Al-K$\alpha$ and hard Cr-K$\alpha$ sources are provided in the supplementary information.

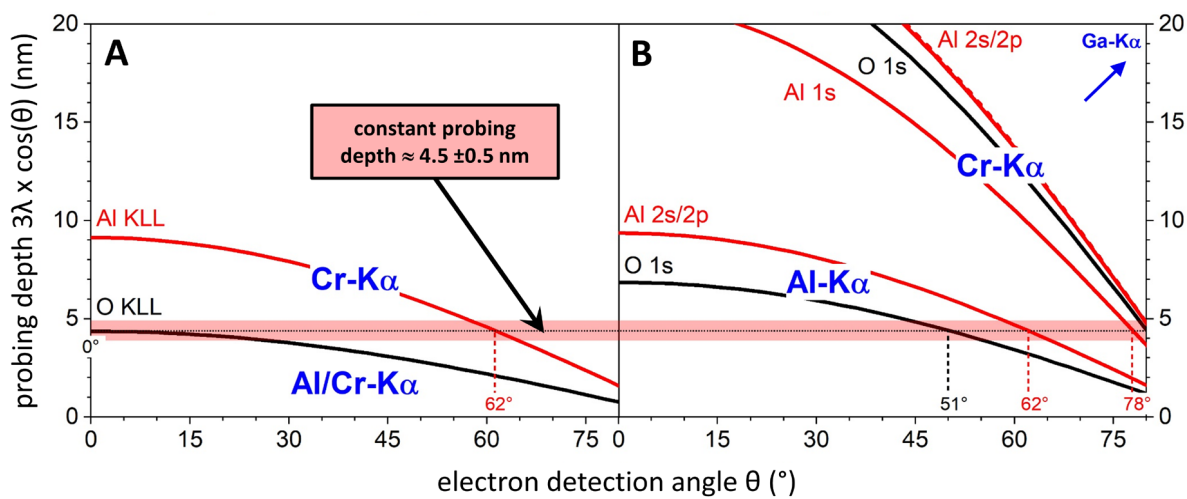

**Figure 3:** Angle-dependent probing depths of the Al KLL and O KLL Auger electron lines (A), as well as the corresponding Al 2s, Al 2p and O 1s core level emission lines (B), in amorphous $Al_2O_3$ for soft Al-K$\alpha$ and hard Cr-K$\alpha$ X-ray sources. The detection angle is given relative to the specimen surface normal. The inelastic mean free paths ($\lambda$) for determination of the probing depths were calculated from the well-known TTP-2 model (see Experimental Section). It follows that the chemical state analysis of Al cations and O anions in $Al_2O_3$ can be performed at constant depth by combining soft and hard X-rays with specific detection angles for each line. Determining the Auger parameter of Al or O only with the hard X-ray source results in large differences in probing depth.

It can thus be concluded that chemical state AP analysis by hard X-rays generally results in very different probing depths of the studied core photoelectron and core-core Auger lines. This might be unproblematic for chemical state AP studies of homogenous bulk compounds, but becomes very critical for studying the local chemical states of atoms constituents in inhomogeneous surface regions. For example, AP analysis of a thin film using a given hard X-ray energy becomes erroneous if the film experiences depth-dependent gradients in the composition, structure or density, or if depth-dependent shifts in the Fermi level occur due to differential charging and/or band bending (see below). Since the kinetic energy of an Auger electron is independent of the incident X-ray energy, the probing depth of the core-core Auger line can only be varied by varying the electron detection angle. The probing depth of the core photoelectron line can be tuned to match the probing depth of the core-core Auger line by variation of the X-ray excitation energy and/or the detection angle. This implies that state-of-the-art chemical state AP studies at constant probing depth in a laboratory environment requires at least the combination of a soft and hard X-ray source. As such, the core-core Auger lines can be excited using the hard X-ray source, whereas the core photo-electron lines can be probed by either the soft or the hard X-ray to closely match the probing depth of the Auger line. By selecting the appropriate excitation source and electron detection angle for each line, the AP analysis can be performed at approximately constant probing depth. In the case of $Al_2O_3$, the measurement conditions would be Al KLL with Cr-K$\alpha$ at $\theta \approx 62°$, Al 1s with Cr-K$\alpha$ at $\theta \approx 78°$, Al 2s/2p with Al-K$\alpha$ at $\theta \approx 62°$, which results in an approximate constant probing depth of $4.5 \pm 0.5$ nm (see Fig. 3).

A first case study on the AP analysis of oxides using a lab-based dual-source XPS/HAXPES system was performed at the research laboratories of Ulvac-Phi in Japan using the commercial PHI Quantes XPS/HAXPES system of Physical Electronics. The PHI Quantes system is equipped with scanning monochromatic Al-K$\alpha$ and Cr-K$\alpha$ X-ray sources. As part of our test measurement schedule, we performed an AP analysis of the local chemical states of Al and O in an amorphous $Al_2O_3$ (am-$Al_2O_3$) layer, as produced by atomic layer deposition (ALD), in comparison to the corresponding local chemical states in a single-crystalline sapphire ($\alpha$-$Al_2O_3$) reference. Given the limited time frame of this preliminary study, all spectra were recorded at a normal detection angle of 0° only (i.e. normal to the sample surface). AP analysis on the basis of the O and Al KLL Auger lines, as measured with Cr-K$\alpha$, and the O 1s and Al 2p photoelectron lines, as recorded with Al-K$\alpha$, then results in an approximate probing depth of about 7.5 ± 2.5 nm (see Fig. 4). The change of the modified Al and O Auger parameters can be visualized by constructing a so-called Wagner (or chemical state) plot by plotting the kinetic energy of the Auger line as a function of the (reverse) binding energy of the photoelectron line. The lines with slope −1 in the chemical-state plot then represent a constant value of the modified Auger parameter: see Figs. 4a and b. Corresponding reference values of $\alpha'_{Al}$ and $\alpha'_{O}$ for sapphire, as reported for non-monochromated Al-K$\alpha$ or Mg-K$\alpha$ sources (corresponding to a very similar information depth),[37] are also indicated by the dashed lines.

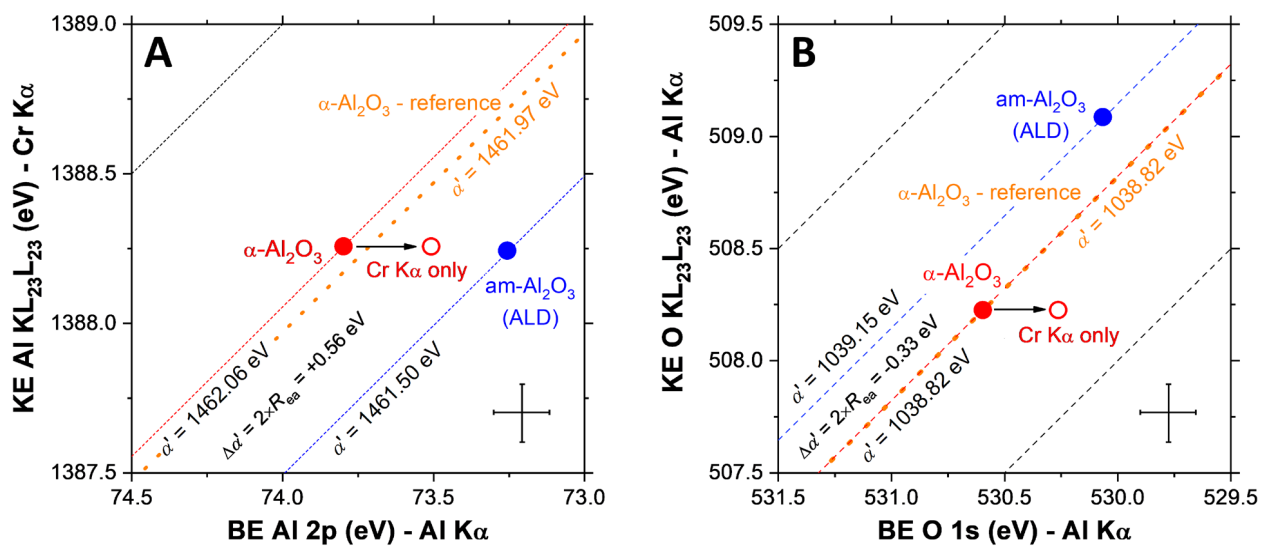

**Figure 4:** Auger parameter analysis (Wagner or chemical state plot) of the local chemical states of Al cations and O anions in an amorphous $Al_2O_3$ (am-$Al_2O_3$) layer, as produced by atomic layer deposition (ALD), in comparison to the corresponding local chemical states in a single-crystalline sapphire ($\alpha$-$Al_2O_3$) reference. The AP analysis was performed using a Cr-K$\alpha$ X-rays for measuring the Al KLL line and Al-K$\alpha$ X-rays for measuring the Al 2p, O 1s and O KLL lines, resulting in an approximate probing depth of 7.5 ± 2.5 nm. A) Shows the AP for Al, B) shows the AP for O. The results for $\alpha$-$Al_2O_3$ are in excellent agreement with previous values from literature.[37] When using only Cr-K$\alpha$ radiation, the Auger and photoelectron lines originate from different depth resulting in an apparent shift of the Auger parameter.

As follows from Figs. 4a and b, the AP values of Al cations and O anions in $\alpha$-$Al_2O_3$ by dual-source XPS/HAXPES are in excellent agreement with previous values obtained by non-monochromated soft X-rays. However, the local chemical states of Al and O in the am-$Al_2O_3$ ALD layer distinctly differ from those in $\alpha$-Al$_2$O$_3$, which can be attributed to the lower density and different local coordination spheres in am-Al$_2$O$_3$ as compared to $\alpha$-Al$_2$O$_3$.[26,29–34] Namely, besides the lower density of the am-Al$_2$O$_3$ ALD layer due to the presence of free volume, the interstitial Al cations in $\alpha$-Al$_2$O$_3$ are all octahedrally coordinated by O, whereas am-Al$_2$O$_3$ phases typically have a mixed coordination of tetrahedrally and octahedrally coordinated Al,[32,33] and even penta-coordinated Al,[39] depending on the synthesis method and processing history. This results in a more pronounced (positive) shift of the AP of the Al cations in the am-Al$_2$O$_3$ ALD layer with respect to the $\alpha$-Al$_2$O$_3$ reference (i.e. $\Delta\alpha'_{Al}$ = 0.56 eV) as compared to the (negative) shift of the corresponding O AP (i.e. $\Delta\alpha'_{O}$ = 0.33 eV). In a future research work, the ALD synthesis conditions of the defective am-Al$_2$O$_3$ ALD layers could be systematically tuned such that their AP values approach those of $\alpha$-Al$_2$O$_3$, thereby likely also achieving similar properties.

As reflected in Fig. 3, the Al 2p core photoelectron line recorded by hard Cr-K$\alpha$ X-rays has a much larger probing depth than the respective core-core Auger line. To investigate the effect of the choice of the X-ray excitation source on the AP analysis of a supposedly homogenous bulk compound, the reference values of $\alpha'_{Al}$ and $\alpha'_{O}$ for $\alpha$-Al$_2$O$_3$ were also determined by using the photoelectron and Auger lines excited by hard Cr-K$\alpha$ X-rays only: see open markers in Figs. 4a and b, respectively. Surprisingly, even for the supposedly homogeneous $\alpha$-Al$_2$O$_3$ reference, the AP parameters values of Al and O are both shifted to lower AP values due to a selective and equal shift of the Al 2p and O 1s photoelectron lines towards higher binding energies (as compared to the respective Al-K$\alpha$ photoelectron lines). The $\alpha$-Al$_2$O$_3$ reference is highly insulating and therefore all measurements related to Fig. 4 were performed by electrically decoupling the insulating samples from ground, while applying dual beam charge neutralization using a combination of low energy electrons and Ar$^+$ ions to achieve steady-state surface charge compensation. However, apparently the floating $\alpha$-Al$_2$O$_3$ specimen exhibits differential charging as a function of depth below the surface.[40] Notably, a X-ray induced surface photo-voltage or junction-voltage in photoactive materials[19,41] can give rise to similar artifacts. This will be unproblematic for chemical state AP parameter analysis at constant depth using the combination soft and hard X-rays (see above), but will produce falsified results if the core photoelectron and Auger lines are probed at different depths with the hard X-ray source only. If only hard X-ray radiation is available, this effect can be remedied by setting the detection angle for the photoelectron line measurements to grazing angles (> 75°) to achieve a near constant probing depth for both photoelectron and Auger electron lines (see Fig. 3).

### 3. Calibration of the energy scale:

In-depth chemical state AP analysis also requires an accurate calibration of the linearity of the kinetic / binding energy scale of the analyzer over the full kinetic energy range. A single point calibration, such as referencing only the Au 4f$_{7/2}$ line, is insufficient to ensure good linearity of the energy scale. For a more accurate calibration of the analyzer, a multipoint calibration using well-defined reference samples has to be performed. In conventional XPS measurements, the binding energy positions of the Au 4f$_{7/2}$, Ag 3d$_{5/2}$ and Cu 2p$_{3/2}$ photoemission lines are well-defined (ISO #15472; see Table 1) and provide a sufficient spread in kinetic energy to perform an accurate energy scale calibration up to binding energies of about 1200 eV (Table 1). However, for modern lab-based HAXPES instruments, the binding energy range exceeds this range by far and the binding energies of deeper core levels, which could provide additional reference points towards higher binding energies (i.e. low kinetic energies), are not nearly as well defined. As an alternative, we suggest measuring a well-established Auger emission lines of standard reference samples (e.g. metallic Cu and Ag) at low kinetic (apparent high binding-) energies such as the Cu $L_3M_{45}M_{45}$ of pure Cu or the Ag $M_4N_{45}N_{45}$ line of pure Ag to perform a multipoint calibration of the energy scale on a single-source lab-based HAXPES system. For a DS-XPS spectrometer, the same reference sample measured with the soft and hard X-ray source already provides two sets of data points at opposite sides of the kinetic energy scale, which can be straightforwardly used for an effective two-point calibration of the energy scale (e.g. Ag $3d_{5/2}$ offers two reference points at 1118.5 eV and 5046.5 eV, in a spectrometer utilizing Al-Kα and Cr-Kα radiation). To offer additional data points for calibration of even wider energy scales, we also report the thus-calibrated positions of the Ag $2p_{3/2}$, Au $3d_{5/2}$ and Ag 2s lines (see Table 1).

**Table 1:** Suggested reference photoelectron and Auger lines for calibration of lab-based XPS/HAXPES instruments. Shown are the binding energy (BE) and kinetic energies (KE) for selected lines of Au, Ag and Cu. The different excitation energies in dual-source instruments allow for a multipoint calibration of the kinetic energy scale using well-established reference samples. *lines used for calibration in this work.

| Feature | BE /eV | $KE_{Al-Kα}$ /eV | $KE_{Ag-Lα}$ /eV | $KE_{Cr-Kα}$ /eV | $KE_{Ga-Kα}$ /eV | Ref. |
|---|---|---|---|---|---|---|
| Fermi Edge | 0 | 1486.7 | 2948.3 | 5414.7 | 9251.7 | |
| Au $4f_{7/2}$ | 83.96 ±0.02 | 1402.74* | 2864.34 | 5330.74* | 9167.74 | ISO #15472 |
| Ag $3d_{5/2}$ | 368.21 ±0.02 | 1118.49* | 2580.09 | 5046.49* | 8883.49 | ISO #15472 |
| Cu $2p_{3/2}$ | 932.62 ±0.02 | 554.08 | 2015.68 | 4482.08 | 8319.08 | ISO #15472 |
| Au $3d_{5/2}$ | 2206.7 ±0.1 | - | 741.6 | 3208 | 7045 | this work |
| Ag $2p_{3/2}$ | 3352.7 ±0.1 | - | - | 2062 | 5899 | this work |
| Ag 2s | 3807.7 ±0.1 | - | - | 1607 | 5444 | this work |
| Cu $L_3M_{45}M_{45}$ | - | 918.62 ±0.03 | 918.62 ±0.03 | 918.62 ±0.03 | 918.62 ±0.03 | 42,43 |
| Ag $M_4 N_{45}N_{45}$ | - | 357.81 ±0.03 | 357.81 ±0.03 | 357.81 ±0.03 | 357.81 ±0.03 | 42 |

The combination of hard and soft X-rays by DS-XPS for chemical state AP analysis of insulating materials poses an additional challenge. The different penetration depths and incident photon fluxes of the soft and hard X-ray sources may result in changes in differential charging of the specimen surface[40] when switching from soft to hard X-rays during the dual-source analysis, even if charge neutralization is applied (see Figure 1b). Analogously, Fermi level shifts by X-ray induced surface photo-voltage or junction-voltage in the case of photoactive materials should depend on the penetration depth and flux of the X-rays.[19,41] Since a common Auger line in the recorded soft and hard X-ray spectra by DS-XPS always originates from the same probing depth, it is proposed here to align the absolute kinetic energy scales of the recorded soft and hard X-ray spectra on a common Auger line of the studied compound. As such, differences in depth-dependent shifts of the recorded photoelectron and Auger lines due to electrical charging or X-ray induced Fermi-level shifts between the soft or hard X-ray source can be largely cancelled out. Figure 5 illustrates the concept of the energy scale alignment using the O KLL of $Al_2O_3$ recorded by DS-XPS using Al-Kα and Cr-Kα sources.

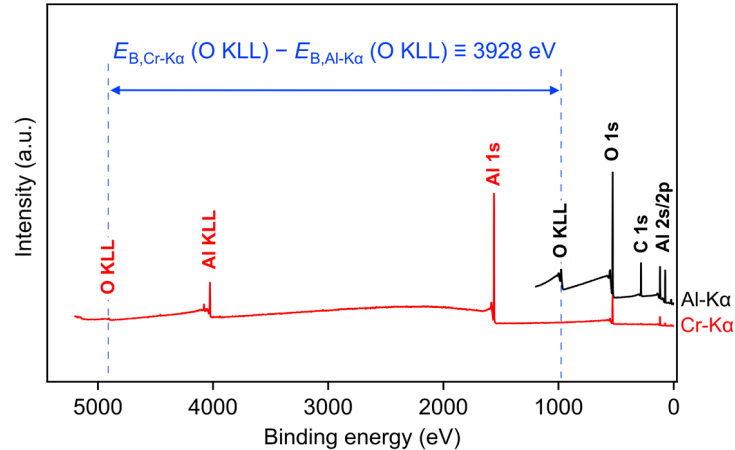

**Figure 5:** The combination of multiple excitation sources requires careful calibration and alignment of the energy scales. The absolute energy scale alignment for different X-ray excitation sources can be performed by aligning a common Auger line with constant probing depth, such as the O KLL line for oxide compounds, at a defined reference value.

## 4. Summary and future directions

The development of state-of-the-art lab-based HAXPES spectrometers offers exciting new possibilities for materials science at surfaces and interfaces, which complement standard XPS measurements. With the increase in information-depth, new experimental challenges and potential pitfalls arise. When different photoelectron and Auger electron lines are combined for local chemical state analysis, the probing depth has to be considered. We suggest a straightforward measurement principle for studying the Auger parameter at near-constant information depth, which is based on angle-resolved measurements using a combination of soft and hard X-ray sources. By following this measurement strategy, previously well-established reference values for bulk α-$Al_2O_3$ were accurately reproduced using a modern dual-source XPS/HAXPES (DS-XPS) instrument. In contrast, a conventional AP analysis using only the hard X-ray source on the same instrument leads to an artificial shift in the Al and O Auger parameters due to a large spread in probing depths of the recorded core photoelectron (O 1s, O 1s) and corresponding KLL Auger lines (O KLL and Al KLL), which confirms the requirement to consider the different probing depths of the employed photoelectron and Auger lines for state-of-the-art chemical state analysis.

State-of-the-art chemical-state studies of functional materials by lab-based DS-XPS profits from an integrated system with combined synthesis, processing and environmental exposure, as will be implemented at the Empa laboratories in 2020. This enables diverse scientific studies on a day-to-day basis to reveal the evolution of the chemistry and electronic structure of buried interfaces in thin films, catalysts and other types of functional (nano-)materials during successive synthesis, processing and environmental exposure steps.

## 5. Methods

The data reported in this work was collected at the ULVAC-Phi facility in Chigasaki, Japan. Hard X-ray photoelectron spectroscopy was performed using a Physical Electronics Quantes spectrometer featuring monochromated Al-K$\alpha$ as well as monochromated Cr-K$\alpha$ radiation at energies of 1486.7 eV and 5414.7 eV, respectively. Unless otherwise specified the analysis was conducted at a pass energy of 69 eV and an electron take of angle was 0° relative to the substrate normal to maximize the count rates. Charge neutralization was achieved using a low-energy electron flood gun. The energy scale of the instrument was calibrated using an effective two-point calibration using Au $4f_{7/2}$, Ag $3d_{5/2}$ and Cu $2p_{3/2}$ photoemission lines excited using both sources to create reference points at both ends of the kinetic energy scale (see Table 1 for details). For the Auger parameter study the different static charges for the Cr-K$\alpha$ measurements and the Al-K$\alpha$ measurements were compensated by aligning the O KLL line (and shifting the Cr-K$\alpha$ related spectra accordingly). The amorphous (am) $Al_2O_3$ sample was deposited via atomic layer deposition (ALD) in an Ultratech Fiji G2 plasma-enhanced ALD system at 150 °C in thermal mode from water and trimethylaluminum (TMA) precursors. A commercial sapphire substrate (Crystec) with (0001) cut was used as $\alpha$-$Al_2O_3$ reference. The samples were mounted using non-conductive adhesive tape to avoid differential charging. The inelastic mean free paths ($\lambda$) for determination of the angle-dependent probing depths in am-$Al_2O_3$, am-$SiO_2$ and $TiO_2$ were calculated from the well-known TTP-2 model, while assuming oxide densities of 3.6, 3.8 and 2.2 g/cm$^3$ and band gaps of 6 eV, 3.2 eV and 6 eV for am-$Al_2O_3$, am-$SiO_2$ and $TiO_2$, respectively.

**Acknowledgements**

The authors would like to thank ULVAC-PHI for the opportunity to conduct the test measurements. We acknowledge financial support from the Swiss National Science Foundation (R'Equip program, Proposal No. 206021_182987). The Laboratory for Thin Films and Photovoltaics at Empa is acknowledged for providing the am-$Al_2O_3$ sample.

**Supporting information:**

**Advanced chemical state studies of oxide films by lab-based HAXPES combining soft and hard X-ray sources**

S. Siol[1*], J. Mann[2], J. Newman[2], T. Miyayama[3], K. Watanabe[3], P. Schmutz[1], C. Cancellieri[1], L. P. H. Jeurgens[1*]

[1] Empa, Swiss Federal Laboratories for Materials Science and Technology,
 Laboratory for Joining Technologies and Corrosion, Dübendorf. Switzerland

[2] Physical Electronics, 18725 Lake Drive East, Chanhassen, MN 55317, USA

[3] Ulvac-PHI, 2500 Hagisono, Chigasaki, Kanagawa, 253-8522, Japan

**E-Mail:**
Lars.Jeurgens@empa.ch
Sebastian.Siol@empa.ch

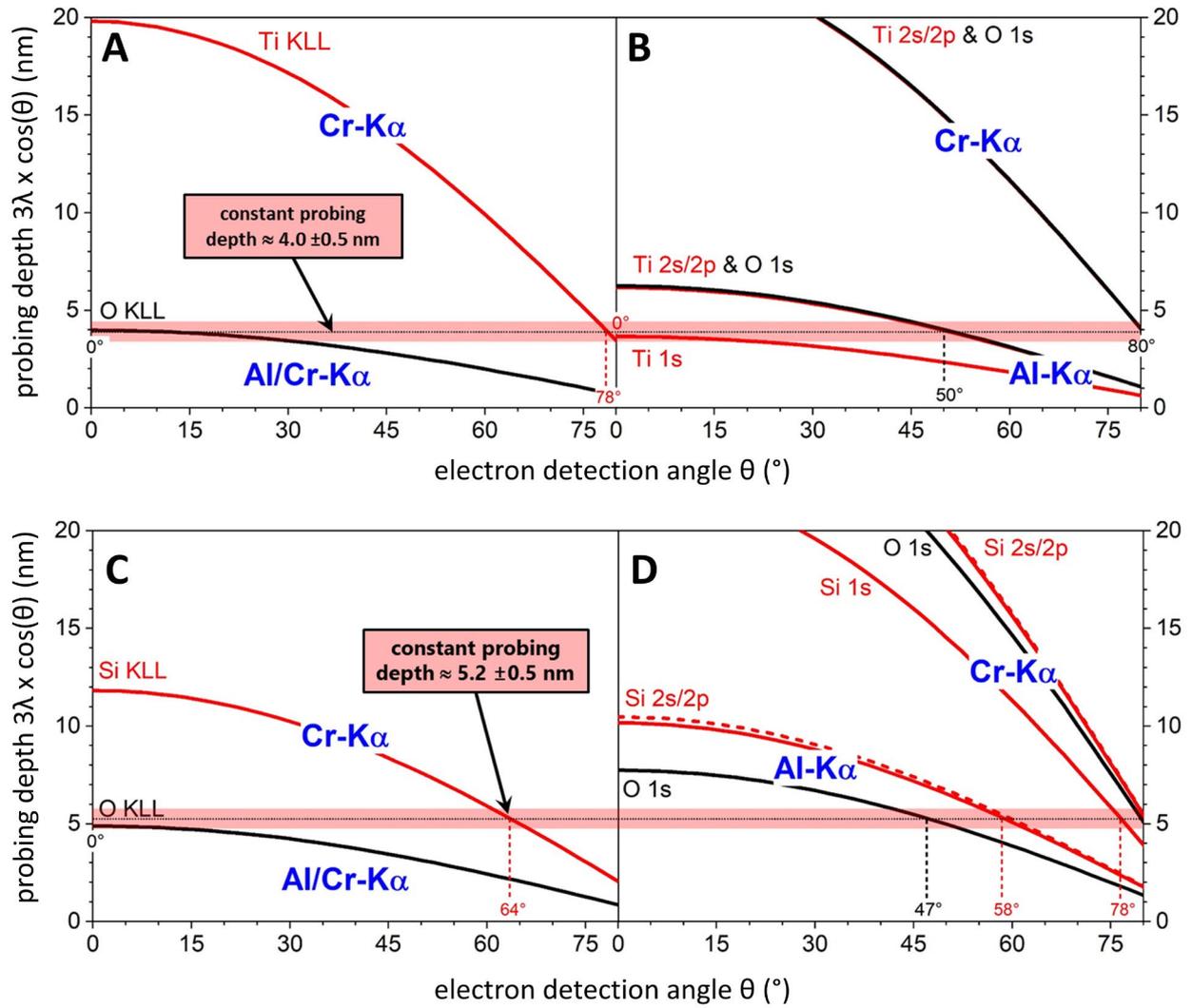

**Figure S1:** Angle-dependent probing depths for Auger electron lines and the corresponding photoelectron lines for soft Al-Kα and hard Cr-Kα X-ray sources; A & B show the relevant lines for chemical state analysis of titanium oxide, C & D show the relevant lines for silicon oxide. The detection angle is given relative to the specimen surface normal. The inelastic mean free paths (λ) for determination of the probing depths were calculated from the well-known TTP-2 model, while assuming typical densities and band gaps for amorphous Ti and Si oxides.